\documentclass[letterpaper, 10 pt, conference]{ieeeconf}

\IEEEoverridecommandlockouts                              
\usepackage[utf8]{inputenc}
\usepackage[T1]{fontenc}
\usepackage{graphicx}
\usepackage{caption}
\usepackage{hyperref}
\usepackage{cite}
\usepackage{rotating}
\usepackage{amsmath}
\usepackage{amsfonts}
\usepackage{pdflscape}
\usepackage{afterpage}
\usepackage{lscape}
\usepackage{array}
\usepackage{fancyhdr}

\title{\LARGE \bf
Style transfer as data augmentation: evaluating unpaired image-to-image translation models in mammography
}

\author{Emir Ahmed$^{1}$ and Spencer A. Thomas$^{1}$ and Ciaran Bench$^{1}$
\thanks{$^{1}$E. Ahmed, S. Thomas, and C. Bench are with the Department of Data Science and AI,
        National Physical Laboratory, Hampton Road, Teddington, United Kingdom, TW11 0LW.
        {\tt\small emir.ahmed@npl.co.uk}}%
}

\begin{document}

\maketitle
\thispagestyle{empty}
\pagestyle{empty}

\thispagestyle{fancy}
\fancyhf{} 
\fancyfoot[L]{\small This work has been submitted to the IEEE for possible publication. Copyright may be transferred without notice, after which this version may no longer be accessible.} 
\renewcommand{\headrulewidth}{0pt} 
\renewcommand{\footrulewidth}{0pt} 

\begin{abstract}

Several studies indicate that deep learning models can learn to detect breast cancer from mammograms (X-ray images of the breasts). However, challenges with overfitting and poor generalisability prevent their routine use in the clinic. Models trained on data from one patient population may not perform well on another due to differences in their data domains, emerging due to variations in scanning technology or patient characteristics. Data augmentation techniques can be used to improve generalisability by expanding the diversity of feature representations in the training data by altering existing examples. Image-to-image translation models are one approach capable of imposing the characteristic feature representations (i.e. style) of images from one dataset onto another. However, evaluating model performance is non-trivial, particularly in the absence of ground truths  (a common reality in medical imaging). Here, we describe some key aspects that should be considered when evaluating style transfer algorithms, highlighting the advantages and disadvantages of popular metrics, and important factors to be mindful of when implementing them in practice.  We consider two types of generative models: a cycle-consistent generative adversarial network (CycleGAN) and a diffusion-based SynDiff model. We learn unpaired image-to-image translation across three mammography datasets. We highlight that undesirable aspects of model performance may determine the suitability of some metrics, and also provide some analysis indicating the extent to which various metrics assess unique aspects of model performance. We emphasise the need to use several metrics for a comprehensive assessment of model performance.
\newline

\indent \textit{Clinical relevance}— Image-to-image translation models are used to augment the training sets of disease classifiers, which can supplement small datasets and potentially reduce bias. This can improve model generalisability, equitability and performance, and hence, patient outcomes. This work describes important factors that need to be considered to effectively evaluate their performance. 

\end{abstract}

\section{Introduction}
\label{sec:intro}
Mammography is a widely used X-ray breast imaging technique that helps facilitate the early detection of cancer. One in eight women in the United States will develop breast cancer at some point during their lifetime, where  screening programs have drastically improved patient outcomes \cite{bevers2009breast}. Yet the need for human clinicians to manually inspect images places a heavy burden on healthcare systems. Automated diagnostic protocols based on deep neural networks have the potential to assist clinicians in making their diagnoses more efficiently and accurately \cite{wang2024mammography}. However, poor generalisability to unseen data is a major barrier to routine application. For example, a deep learning model trained to classify whether a patient has breast cancer may not perform as well when applied to mammograms from different scanners \cite{mckinney2020international}. This is because of differences in the data domains describing either set of images. A data domain $\mathcal{D} = \{\chi, P(x), P(x,y)\}$ is composed of an input feature space $\chi$ (a vector space containing all image features), a marginal distribution $P(x)$, and a joint probability distribution $P(x,y)$. Here, $x$ is a member of the set $X$ of $N$ training example inputs $x_1, x_2, ... x_N \in X$ and $y$ is a member of the corresponding set of ground truths $y_1, y_2, ...y_N \in Y$. A significant discrepancy between $\mathcal{D}_{\text{train}}$ and $\mathcal{D}_{\text{test}}$ results in poor generalisability of models trained on data from $\mathcal{D}_{\text{train}}$ when applied to data from $\mathcal{D}_{\text{test}}$.

Expanding the training set with additional/diverse examples can help align the two domains. However, collating a large set of mammograms from different sources is challenging due to patient privacy laws and complications with coordinating data collection efforts across institutions. Instead, data augmentation schemes can be used to modify existing datasets to produce additional training examples. These modifications often take the form of generic geometric transformations, such as flipping and warping \cite{costa2019data}. However, more complex and domain-specific transformations that align source feature representations with those of the target domain can be applied using certain types of deep neural networks. This task is referred to as image-to-image translation, performed by neural style transfer models that adapt source images to appear more like images from the target domain \cite{jing2019neural}. These models have been used in mammography studies, where the inclusion of augmented examples has improved the performance of classification models \cite{wang2020mr}. With that said, style transfer models are known to suffer from overfitting and other undesirable behaviours, motivating a need to carefully evaluate their performance. 

There are two key concepts to assess: the degree to which the target style (i.e. the feature representations for common objects) has been imposed onto the source domain, and whether the transformed/adapted image is still representative of the original tissue \cite{ozbey2023unsupervised, ding2020image}. The ideal behaviour of a given model is to make it appear as though the image of a given source domain tissue has been acquired using the same technology used to acquire images from the target domain. However, when considering images acquired from different patient populations, differences in tissue composition (e.g. density) may be imposed on the source domain images in addition to the stylistic features imposed by the properties of the different scanners. This assessment is particularly challenging in the absence of ground truths (e.g. the same tissue images with a different scanner), requiring the use of higher-order statistical measures of stylistic content.  While several metrics have been proposed, none provide a fully comprehensive assessment of model performance \cite{bench2025trustworthyimagetoimagetranslationevaluating}.

\section{Methods}
\label{sec:methods}

\subsection{Outline}
We conduct experiments that exhibit the use of various metrics and highlight important aspects to consider when evaluating the performance of style transfer models. We employ the two most common frameworks for generative modelling by using a cycle-consistent generative adversarial network (CycleGAN) \cite{zhu2017unpaired} and a diffusion-based SynDiff model \cite{ozbey2023unsupervised}. We train these to perform unpaired image-to-image translation on patches parsed from images from three open source mammography datasets: VinDr Mammo (VDM) \cite{nguyen2023vindr}, the Chinese Mammography Database (CMMD) \cite{cai2023online}, and the Curated Breast Imaging Subset of the Digital Database for Screening Mammography (CBIS-DDSM) \cite{sawyer2016curated}.

\subsection{Evaluating style}
Differences in the styles of two image sets is typically assessed by observing the similarities between the distribution of features found in adapted source and target domain images (i.e. how often particular features are found in either dataset). Manual feature extraction-based methods, such as gray-level co-occurrence matrices and gabor filters, can be used to acquire features contained in the images. However, these do not necessarily capture the full extent of features that define the style of a domain. Networks trained to perform tasks that require the encoding of more complex feature representations (e.g. classification networks) may be preferable. 

The Fr\'{e}chet Inception Distance (FID)  \cite{heusel2017gans} is one popular metric that implements the Inception-V3 architecture to extract activations from images drawn from a given domain (e.g. the adapted source images $\hat{x}_1, \hat{x}_2, ..., \hat{x}_N \in \hat{X}$). The mean $\mu_{\hat{x}}$ and covariance $\Sigma_{\hat{x}}$ of the resultant activations are used to parameterise a multivariate Gaussian distribution $\mathcal{N}_{{\hat{x}}}(\mu_{\hat{x}}, \Sigma_{\hat{x}})$, that is then compared to the corresponding distribution computed from a set of target images (unpaired) $\mathcal{N}_{y}(\mu_y, \Sigma_y)$ with a Fr\'{e}chet distance:
\begin{equation}   
\begin{split}
        \mathrm{FID}(\hat{X},Y) = d_F\left(\mathcal{N}_{{\hat{x}}}(\mu_{\hat{x}}, \Sigma_{\hat{x}}), \mathcal{N}_{y}\left(\mu_y, \Sigma_y\right)\right)^2= \\
        \left\|\mu_{\hat{x}}-\mu_y\right\|_2^2+\text{tr}\left(\Sigma_{\hat{x}}+\Sigma_y-2\left(\Sigma_{\hat{x}} \Sigma_y\right)^{\frac{1}{2}}\right).
    \end{split}
\end{equation}

However, modelling the distribution as a multivariate Gaussian is a strong assumption to make about the distributions of activations, especially considering the representations are produced from layers with ReLU activations, whose distribution is inherently unsmooth. The authors of \cite{binkowski2018demystifying} note that typically 2\% of Inception representations were zero, indicating a complex, non-Gaussian form for the distribution of activations in their experiments.

The Kernel Inception Distance (KID) \cite{binkowski2018demystifying} is an alternative metric that measures the squared maximum mean discrepancy between activation distributions. Here, a polynomial kernel is used to assess similarities of activations for the different images, 
\begin{equation}
k(\alpha, \omega)=\left(\frac{1}{d} \alpha^{\top} \omega+1\right)^3,
\label{eq:kernel}
\end{equation}
where $\alpha$ and $\omega$ are activations produced from two inputs and $d$ is their dimensionality. Here, we represent the application of this kernel onto the activations produced using two different images, e.g., $x_1$ and $\hat{x}_1$, with $k(\phi(x_1),\phi(\hat{x}_1))$, where $\phi$ represents application of Inception-V3 to acquire activations.  The KID uses the polynomial kernel to assess the similarity of the adapted source images with themselves $k(\phi(\hat{x}),\phi(\hat{x}))$, the target images with themselves $k(\phi(y),\phi(y))$, and between the adapted source and target images $k(\phi(\hat{x}),\phi(y))$ and $k(\phi(y),\phi(\hat{x}))$. The kernel values are acquired and used to compute the KID, where $m$ and $n$ are the total number of images for the adapted source and target, respectively:
\begin{equation}
\begin{split}
    \text {KID}(\hat{X},Y) = & \ \frac{1}{m(m - 1)} \sum_{i \neq j}^{m} k(\phi(\hat{x}_i), \phi(\hat{x}_j)) \\
    & + \frac{1}{n(n - 1)} \sum_{i \neq j}^{n} k(\phi(y_i), \phi(y_j)) \\
    & - \frac{2}{mn} \sum_{i=1}^{m} \sum_{j=1}^{n} k(\phi(\hat{x}_i), \phi(y_j)).
\end{split}
\label{eq:KID}
\end{equation}

Aside from not requiring a Gaussian distribution, another benefit of using the KID is that Eq. \ref{eq:kernel} enables a comparison of the skewness as well as the mean and variance of the activation distributions. The additional moment provides a more thorough assessment of the similarity between the distributions. The KID has been reported to be more effective than the FID on smaller sets of images \cite{binkowski2018demystifying}, which is appealing in the context of mammography given the typical scarcity of available data. However, it is widely known that the Inception-V3 used in FID and KID is pretrained on the ImageNet dataset (image classification), and so is not optimal for extracting features from mammography images. I.e., because it was trained to classify natural images, it is not optimised to detect features that characterise the style of mammography images. Nonetheless, both the FID and KID are widely used in medical imaging studies in the absence of a domain-specific feature extraction network.  Other metrics like the CLIP (Contrastive Language-Image Pre-Training) Maximum Mean Discrepancy (CLIP-MMD) \cite{jayasumana2024rethinking} use CLIP embeddings that may capture a more diverse range of feature representations but ultimately, it is still not specialised to detect information relevant to characterising mammography images. Here, we simply employ the more commonly used KID and FID.

\subsection{Evaluating content preservation}
Assessing stylistic content alone fails to capture another important aspect of model performance: content preservation. We desire translated images to not only contain features characteristic of the target domain, but also still be representative of the original tissue.  The preservation of the image's structural content, or the underlying tissue components and anatomical features, is often used to assess this aspect of model performance. The key challenge is disentangling this lower-order information from the higher-order statistical information related to style \cite{zhang2011fsim}. A hypothetical, ideal metric would return an optimal score when processing two images of the same tissues acquired with different scanners. However, the metrics typically used to assess this aspect of model performance exhibit this behaviour to varying degrees. Here, we consider reference-based metrics (i.e. those that compare the unadapted source image with an adapted version) because these provide a more direct assessment of changes in content post-translation compared to metrics that do not incorporate a comparison with a reference image.

The mean squared error (MSE) between a source image and its adapted counterpart is one crude metric \cite{zhang2018unreasonable}. However, its optimal value occurs when no adaptation is applied to the image. It also fails to consider spatial dependencies/correlations that can be used to effectively define lower-level structural content. The peak signal-to-noise ratio (PSNR) \cite{ozbey2023unsupervised} is another metric that contains an MSE term, and so suffers similar drawbacks.

The Structural Similarity Index Measure (SSIM) is widely used in mammography and other medical imaging studies \cite{7404021}. In contrast to MSE or PSNR, which use pixel-wise comparisons of image amplitude, differences in the contrast and luminance between groups of pixels are used to measure structure. Here, contextual information that is important to defining structural features is considered, making the metric better equipped to detect distortions which typically span regions of pixels \cite{1284395}. With that said, SSIM is known to be highly sensitive to geometric and scale distortions \cite{7404021}.

The Deep Image Structure and Texture Similarity (DISTS) \cite{ding2020image} is another metric, where activations from an image are extracted using a unique variant of the VGG (Visual Geometry Group) model pre-trained to classify ImageNet. Here, an image from the source domain dataset is passed through the network, where activations from several layers are acquired and normalised. Activations are also produced from its corresponding adapted source image. The covariance between the corresponding activations as well as the variance of the activations are used to assess structural preservation, while textural similarity is assessed using the mean of the activations from each image. The metric combines these
measurements from different convolution layers using a
weighted sum. The textural similarity, in part, defines the style of the images, which is an unappealing aspect of using this metric to assess content preservation. Another drawback is, like the FID, the activations are produced from a model trained to classify natural images, and so may not be effective at encoding the structural content of mammography images \cite{deshpande2024report}. 

The Feature Similarity Index for Image Quality Assessment (FSIM) \cite{zhang2011fsim} is another metric that uses differences in the phase congruency and gradient magnitudes between pairs made up of an unadapted image and its adapted counterpart to assess changes in image quality. Phase congruency assesses the significance of local structures using Fourier components (based on the premise that visually discernible features coincide with points where the Fourier waves at different frequencies have congruent phases), while the gradient magnitudes detect variations in contrast important to defining structure. The authors argue that the use of phase congruency as the key feature provides a more effective means of extracting low-level feature information related to encoding structural content. 

\subsection{Practical considerations}
The effectiveness of some metrics may be influenced by processing applied to the output, or artefacts imposed onto the image by the model. E.g. some models may impose offsets on translated images that can affect the use of structural similarity metrics. The subsequent correction of these artefacts using other software packages may impose their own artefacts that should be considered when evaluating model performance.

Some metrics (e.g. the continuous wavelet variant of the SSIM discussed in Section \ref{sec:syndiff_content} \cite{cwssim}) may require images in a particular format, where any conversions may result in a loss of information if not performed carefully (e.g. float32 to unsigned 8-bit integers). Numerical precision may also affect the efficacy of some metrics (e.g. the FID, as will be discussed in Section \ref{sec:numerical_prec}). These factors should be carefully considered when using evaluation metrics, or developing post-processing protocols.

\afterpage{
\begin{landscape}

\begin{table}[h!]
\centering
\caption{Metadata table for datasets. Here, CMMD1 and CMMD2 refer to the two subsets composing the full CMMD datset (no density information provided).} 
\label{tab:Metadata}
 \begin{tabular}{|c|p{6cm}|p{6cm}|p{6cm}|}
  \hline
    & VinDr-Mammo \cite{nguyen2023vindr} & CMMD \cite{cai2023online} & CBIS-DDSM \cite{sawyer2016curated} \\
  \hline
   Patient Population Characteristics &

    \textbf{\underline{Abnormality (rate of findings per 100 images)}}
    \newline
    Mass – \textbf{1,226 (6.130)}
    \newline
    Suspicious Calcification – \textbf{543 (2.715)}

    \vspace{0.2cm}
   \textbf{\underline{Breast Density}}
   
     A (Almost entirely fatty) – \textbf{50 (0.5\%)}
     \newline
     B (Scattered areas of fibro-glandular) – \textbf{954 (9.54\%)}
     \newline
     C (Heterogeneously dense) – \textbf{7,646 (76.46\%)}
     \newline
     D (Extremely dense) – \textbf{1,350 (13.5\%)}

    & \textbf{\underline{Abnormality}}
    \newline
    CMMD1:
    \newline
    Mass – \textbf{726 (65.58\%)}
    \newline
    Calcifications – \textbf{158 (14.27\%)}
    \newline
    Both – \textbf{223 (20.14\%)}
    \vspace{0.2cm}
    \newline
    CMMD2:
    \newline
    Mass – \textbf{417 (55.67\%)}
    \newline
    Calcifications – \textbf{98 (13.08\%)}
    \newline
    Both – \textbf{234 (31.24\%)}
    &  \textbf{\underline{Breast Density}}
    \newline
    Calcification Set (\textbf{753 cases}):
    \newline
    1 (Almost entirely fatty) - \textbf{63 cases}
    \newline
    2 (Scattered areas of fibro-glandular) - \textbf{233 cases}
    \newline
    3 (Heterogeneously dense) - \textbf{249 cases}
    \newline
    4 (Extremely dense) - \textbf{207 cases}
    \vspace{0.2cm}
    \newline
    Mass Set (\textbf{891 cases}):
    \newline
    1 (Almost entirely fatty) - \textbf{149 cases}
    \newline
    2 (Scattered areas of fibro-glandular) - \textbf{399 cases}
    \newline
    3 (Heterogeneously dense) - \textbf{252 cases}
    \newline
    4 (Extremely dense) - \textbf{92 cases}
    \\
   \hline
   Common Artefacts & Stripe artefacts & Substantial motion artefacts & Labels, horizontal and vertical lines \\
   \hline
   Scanner Types & 
   SIEMENS
   \newline
   IMS
   \newline
   PLANMED
   & 
   GE Senographe DS
   \newline
   Siemens Mammomat Inspiration
       & 
       DBA
       \newline
       HOWTEK
       \newline
       LUMYSIS
   \\
   \hline
   Pathology & 
   \textbf{\underline{Breast BI-RADS}}
   \newline
   BI-RADS 1 - \textbf{6,703 (67.03\%)} 
   \newline
   BI-RADS 2 - \textbf{2,338 (23.38\%)}
   \newline
   BI-RADS 3 - \textbf{465 (4.65\%)}
   \newline
   BI-RADS 4 - \textbf{381 (3.81\%)}
   \newline
   BI-RADS 5 - \textbf{113 (1.13\%)}
   & 
   \textbf{\underline{Image Categories}}
   \newline
   CMMD1:
   \newline
   Benign – \textbf{544 (49.14\%)}
   \newline
   Malignant – \textbf{563 (50.86\%)}
   \vspace{0.2cm}
   \newline
   CMMD2:
   \newline
   Benign – \textbf{0 (0\%)}
   \newline
   Malignant – \textbf{749 (100\%)}
   & 
   \textbf{\underline{Categories}}
   \newline
   Calcification Set:
   \newline
   Benign - \textbf{414 cases (664 abnormalities)}
   \newline
   Malignant - \textbf{339 cases (381 abnormalities)}
   \vspace{0.2cm}
   \newline
   Mass Set:
   \newline
   Benign - \textbf{472 cases (522 abnormalities)}
   \newline
   Malignant - \textbf{419 cases (448 abnormalities)}
   \\
   \hline
   Mode of Acquisition & Full field digital mammography (FFDM) & Full field digital mammography (FFDM) &  Screen-film mammography (SFM) \\
   \hline
 \end{tabular}

\end{table}

\begin{figure}[h!]
    \centering
    \begin{minipage}{0.49\columnwidth}
        \centering
        \includegraphics[width=.48\columnwidth]{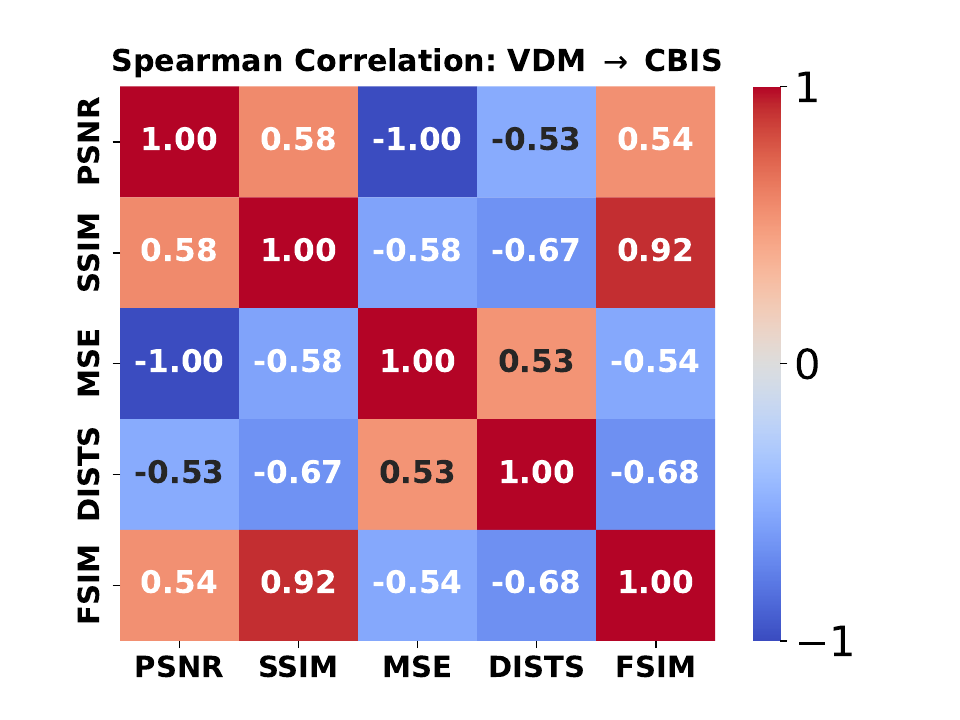}
        \caption{Matrix of Spearman's rank correlation coefficients comparing all the content preservation metrics for the VDM $\rightarrow$ CBIS task}
        \label{fig:correlation_matrix}
    \end{minipage}
    \hfill
    \begin{minipage}{0.49\columnwidth}
        \centering
        \includegraphics[width=.48\columnwidth]{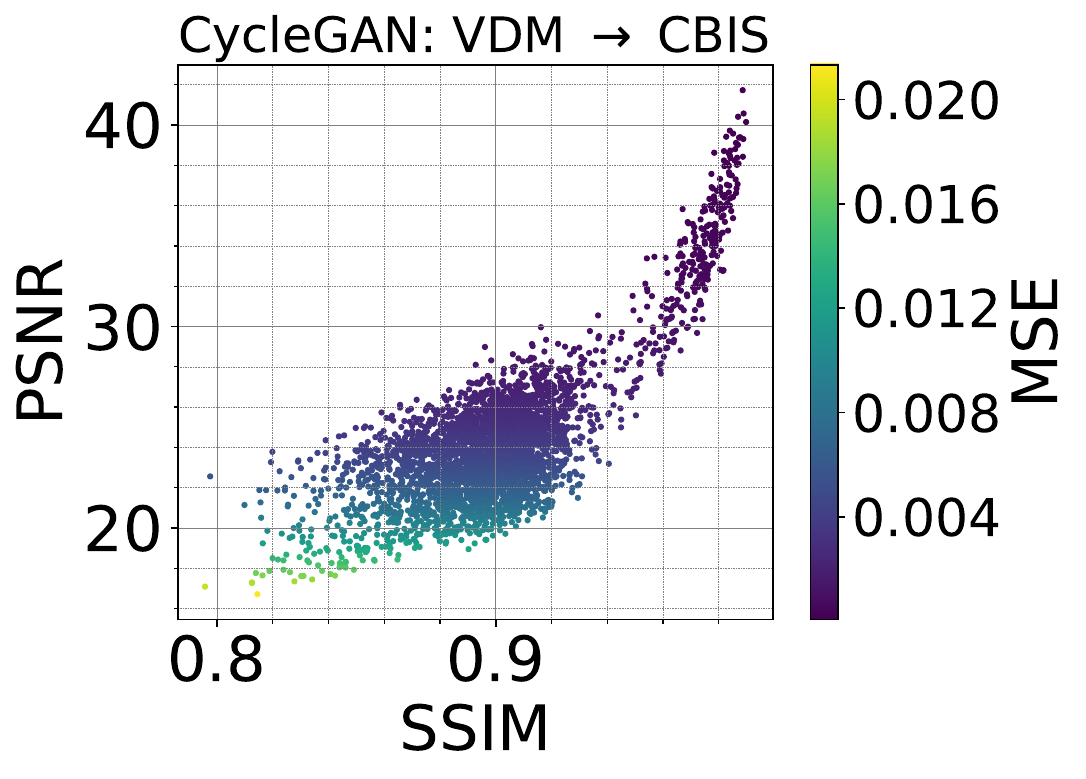}
        \caption{PSNR vs SSIM plot for the VDM $\rightarrow$ CBIS task}
        \label{vindr_cbis_psnr_ssim}
    \end{minipage}%
\end{figure}
\end{landscape}
}

\subsection{Data}

We train a CycleGAN and SynDiff model on 1500 256$\times$256 image patches for each domain for all six tasks (style transfer each way between every pair of the three datasets). We then evaluate 3500 test patches, and apply the metrics discussed in Section \ref{sec:methods}, providing commentary on their use and interpretation.

The characteristics of each dataset are outlined in Table \ref{tab:Metadata}. The variation in the physiological characteristics of the patient populations and in scanning technology provide a significant challenge for the translation models. 

We considered patches parsed from full-scale mammogrpahy images to alleviate issues with memory consumption. These were parsed from whole mammography images in 256$\times$256 pixel patches using a step size of 246 pixels. Only patches containing 99\% of pixels with non-zero values were included in the dataset.

Before decomposing each mammogram into patches, the breast was segmented from the background for VDM and CBIS-DDSM images using Otsu's method \cite{ostu1979threshold} (CMMD was already segmented). We applied contrast inversion to images from all three datasets that had photometric interpretation set to \verb |MONOCHROME1|. We also horizontally flipped all `right' laterality images. Each image was normalised so the maximum pixel value was 1, and the minimum was 0. Padding was applied to images so they had dimensions of 2224$\times$2224 pixels. 
Histogram equalisation was applied to each patch and the values were then normalised. 1500 patches and 3500 patches from each dataset were collated for training and testing, respectively (with no overlap across the training/test sets at the patch or whole image level). The number of training patches was in part determined to accommodate training times with the SynDiff model (i.e. 100 epochs within 4 days)  using the available computational resources (NVIDIA A100 GPU, and an AMD EPYC 7643 2.3 GHz CPU) \cite{bench2025trustworthyimagetoimagetranslationevaluating}.

All adapted images (model outputs) were normalised with pixel values in the range [0,1] before calculating any evaluation metric/subsequent post-processing steps.
\subsection{Training}
We use the default parameterisations of the CycleGAN and SynDiff models, aside from an increased cycle-consistency weight of 1000 for the SynDiff models. Our aim here is not to compare which model is better, but rather to use the results/unique behaviours of each to emphasise important aspects of evaluation. We also use the default number of training epochs associated with each implementation (100 for SynDiff and 200 for CycleGAN), as this was reported to produce sensible results on related experiments \cite{zhu2017unpaired, ozbey2023unsupervised}.

\section{Experiments and Results}

\subsection{Evaluating content preservation (CycleGAN)}

\begin{table}[ht!]
\centering
\caption{Content Preservation Metrics: CycleGAN}
\begin{tabular}{|c|c|c|c|c|c|}
\hline
Task & SSIM & PSNR & MSE & DISTS & FSIM
\\
\hline
CMMD $\rightarrow$ CBIS & 0.86 & 23.7 & 0.0048 & 0.1378 & 0.91 \\
CBIS $\rightarrow$ CMMD  & 0.95 & 33.5 & 0.0016 & 0.0656 & 0.97 \\
\hline
CMMD $\rightarrow$ VDM & 0.96 & 29.5 & 0.0018 & 0.0903 & 0.97 \\
VDM $\rightarrow$ CMMD  & 0.95 & 27.0 & 0.0024 & 0.1333 & 0.96 \\
\hline
VDM $\rightarrow$ CBIS & 0.90 & 24.3 & 0.0046 & 0.1081 & 0.95 \\
CBIS $\rightarrow$ VDM  & 0.96 & 33.0 & 0.0015 & 0.0707 & 0.98 \\
\hline
\end{tabular}
\label{tab:CycleGANrefmerticscores}
\end{table}

We use the scikit-image\footnote{https://scikit-image.org/} package to calculate SSIM, PSNR and MSE,  and other code repositories to calculate DISTS\footnote{https://github.com/dingkeyan93/DISTS} and FSIM\footnote{https://piq.readthedocs.io/en/latest/}. FSIM scores typically range from 0 to 1, where 1 indicates perfect similarity between two images; however values greater than 1 can occur if the adapted source image has greater contrast than its unadapted counterpart. DISTS scores also lie in the range [0,1], but 0 here instead represents perfect similarity. The bounds for SSIM are -1 and 1, where negative structural similarity values correspond to the cases where the local image structures are inverted. PSNR has a lower bound of 0, but does not have an upper bound, making it challenging to assess which score infers a `good quality' of preservation. 

In Table \ref{tab:CycleGANrefmerticscores} we see high scores for SSIM and FSIM, and low scores for MSE and DISTS for all tasks, indicating that much of the structural content of the tissue has been preserved. Fig. \ref{fig:correlation_matrix} displays a correlation matrix between the five metrics for the VDM $\rightarrow$ CBIS task. We choose Spearman's correlation coefficient as we are interested in assessing the monotonicity between pairs of metrics. While a low correlation between two metrics computed for a given test set doesn't necessarily indicate each metric is considering different aspects of image contents, assessing this provides a useful starting point for a deeper analysis of their distinct qualities, and whether in combination they may provide a more comprehensive assessment of model performance. In Fig. \ref{fig:correlation_matrix} we see that PSNR and SSIM have a moderate positive correlation between each other, which broadly agrees with the scatterplot of the two metrics shown in Fig. \ref{vindr_cbis_psnr_ssim}, where the lower cluster of qualitatively less correlated points accompanies a smaller cluster of more highly correlated points. Indeed, this modest correlation is not entirely unexpected given the differences in which structure is assessed with either metric. We see the same score, but negative when comparing SSIM to the MSE, which is expected given the inverse relation between MSE and PSNR. We also see modest correlations when comparing all these metrics with DISTS. In addition to the unique manner with which it acquires information about structure, the metric also encodes information about style/textural features, which likely contributes to the lower correlation with these more structure focused metrics. The FSIM correlates strongly with the SSIM, which could suggest that each ultimately uses similar information when assessing structural content. The variation we observe in the correlations between different metrics highlights that each considers different aspects of image content to assess structure, and that several should be used to provide a more comprehensive assessment of model performance.

\subsection{Evaluating content preservation (SynDiff)}
\label{sec:syndiff_content}

\begin{table}[ht!]
\centering
\label{tab:syndiff_content_preservation}
\caption{Content Preservation Metrics: SynDiff}
\begin{tabular}{|c|c|c|c|c|}
\hline
Task & SSIM & PSNR & CW-SSIM & DISTS
\\
\hline
CMMD $\rightarrow$ CBIS & 0.71 & 21.4 & 0.93 & 0.1427 \\
CBIS $\rightarrow$ CMMD  & 0.75 & 20.9 & 0.92 & 0.1260 \\
\hline
CMMD $\rightarrow$ VDM & 0.57 & 19.1 & 0.96 & 0.1453 \\
VDM $\rightarrow$ CMMD  & 0.59  & 19.4 & 0.95 & 0.1256 \\
\hline
VDM $\rightarrow$ CBIS & 0.52 & 19.6 & 0.94 & 0.1200 \\
CBIS $\rightarrow$ VDM  & 0.54 & 19.7 & 0.94 & 0.1040 \\
\hline
\end{tabular}
\label{tab:SynDiffRefMetricscores}
\end{table}

The SynDiff model was found to produce small offsets of a few pixels to translated images. This resulted in low values for the more generic structural similarity metrics (SSIM and PSNR in Table \ref{tab:SynDiffRefMetricscores}). However, qualitatively, the image contents otherwise appeared to be of high quality. Some variants of common metrics can be more tolerant of small distortions and are better suited to assessing data that exhibits these undesirable model behaviours. We select the continuous wavelet variant of SSIM (CW-SSIM) which has been used in other mammography-based studies\cite{7404021}, and we use DISTS, as in Table \ref{tab:CycleGANrefmerticscores}, which is tolerant to mild geometric translations. Even though SSIM and PSNR are known to be sensitive to such distortions, we retain their scores in our results to show a comparison between sensitive and insensitive metrics. The results in Table \ref{tab:SynDiffRefMetricscores} indicate SSIM and PSNR are heavily affected by the offset imposed by SynDiff, resulting in lower scores compared to CycleGAN. With that said, the more tolerant metrics (CW-SSIM and DISTS), show that a lot of structure is preserved post-adaptation. This is in line with the results achieved with more generic structural similarity metrics on versions of the image where this offset has been corrected (described shortly), indicating that these more tolerant metrics are assessing underlying structure effectively in the face of the offset.

\begin{figure}[ht!]
    \centering
    \includegraphics[width=1\columnwidth]{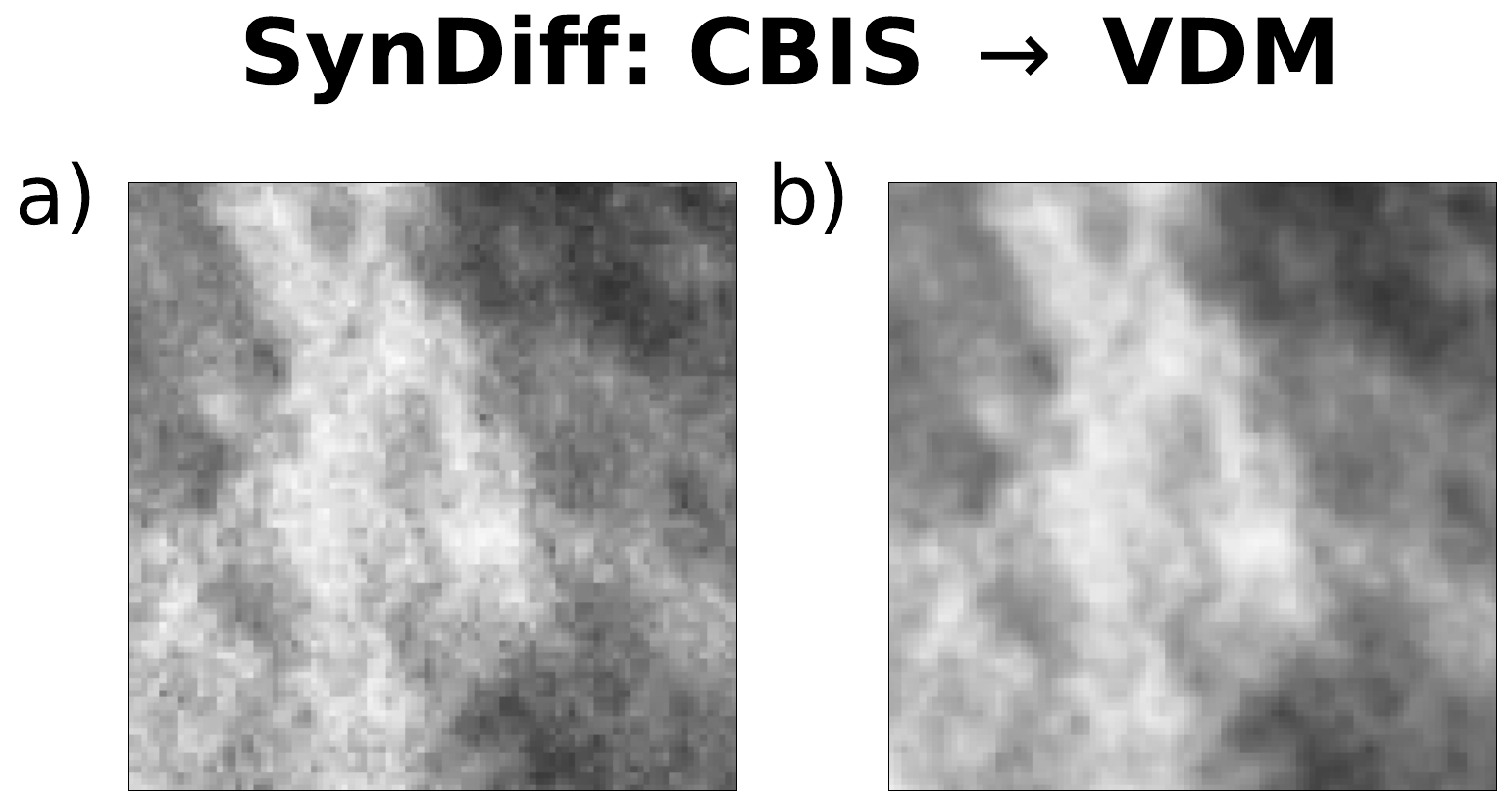}
    \includegraphics[width=.48\columnwidth]{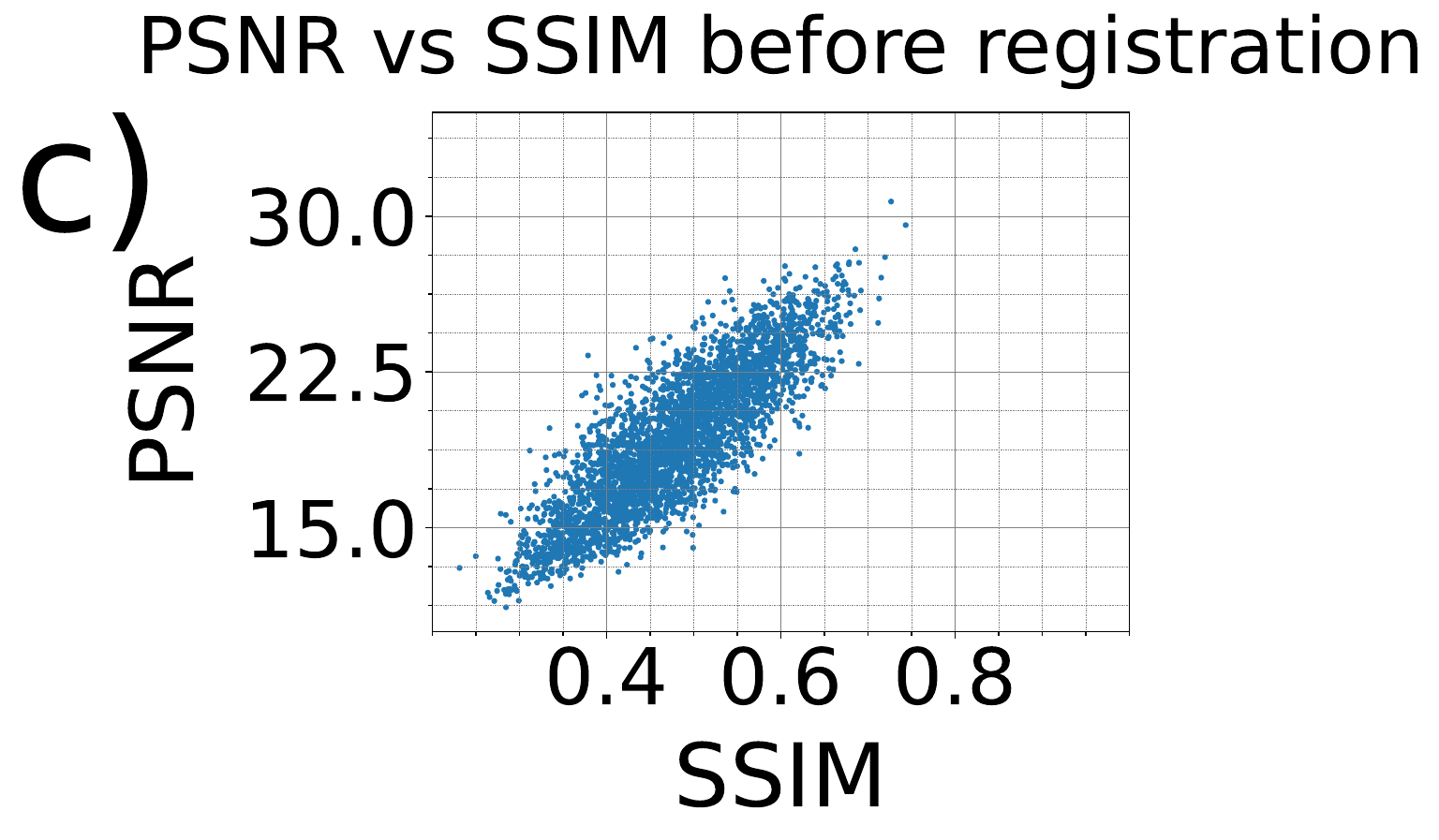}
    \includegraphics[width=.48\columnwidth]{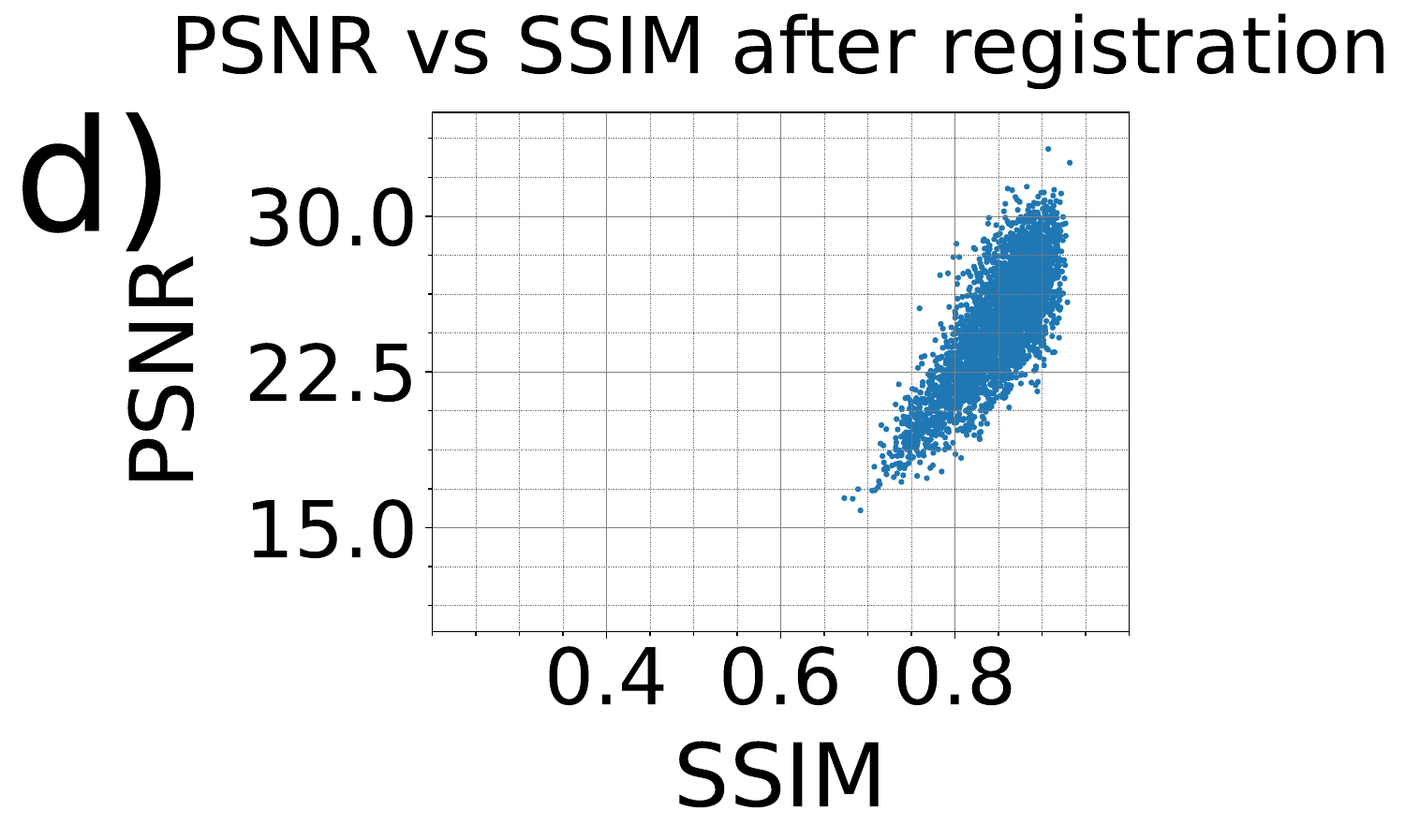}
    \caption{Example image pair of sections of adapted CBIS image patches a) before and b) after registration, for the CBIS $\rightarrow$ VDM task. The adapted CBIS output, after registration, consists of some blur, which could have an impact on SSIM and PSNR scores, c) and d), respectively.}
    \label{fig:Example_CBIS_VINDR}
\end{figure}
We note that in some cases, it may be preferable to correct artefacts like offsets (e.g. to facilitate uncertainty quantification, which may require pixel-wise comparisons of image content) instead of using metrics that tolerate them. However, these can impose their own distortions on image contents, and their effects should be considered when evaluating model performance. 

Here, we demonstrate this approach by correcting the offsets imposed on images by the SynDiff model using a popular processing package, ANTsPy\footnote{https://github.com/ANTsX/ANTsPy}.
As stated previously, we notice that the offset observed differs in magnitude and direction for each SynDiff model. We specifically use the translation-mode functionality from the ANTsPy package, to correct the offsets. We trained a SynDiff model for the CBIS$\leftrightarrow$VDM tasks, distinct to the model used to collate the results in Table \ref{tab:SynDiffRefMetricscores}, and we register the test outputs to their respective source images. To remove any lines of black pixels from the images after registration, we cropped five pixels from each side to ensure that all images had the same dimensions. We notice that post-registered images contain some blurring, such as in Fig. \ref{fig:Example_CBIS_VINDR}, which could have a significant effect when applying evaluation metrics. We also compute the SSIM and PSNR scores after correcting the offsets, hoping to see a better indication of the amount of content preserved, rather than the offset dominating the scores. 

As expected, the SSIM and PSNR scores increase by a large amount after registration (Fig.\ref{fig:Example_CBIS_VINDR}). However the interpretation of these metrics is made more complex given a report that indicates that some metrics, including the SSIM, will produce better scores from blurred images\cite{dohmen2024five}. These results emphasise the need to consider how post-processing can affect image quality and perceptions about model performance.

\subsection{Evaluating style}
We compare the FID and KID values assessed between adapted source images and target images with a baseline score computed from unadapted source images and the target test images.
\begin{table}[ht!]
\centering
\caption{FID scores summary}
\begin{tabular}{|c|c|c|c|}
\hline
Task & Baseline & CycleGAN & SynDiff
\\
\hline
CMMD $\rightarrow$ CBIS & 35.2 & 32.7 & {\bf 15.2}  \\
CBIS $\rightarrow$ CMMD  & 35.2 & 21.4 & {\bf 20.0} \\
\hline
CMMD $\rightarrow$ VDM & 37.2 & {\bf 30.5} & 34.2 \\
VDM $\rightarrow$ CMMD  & 37.2 & {\bf 23.2} & 24.1  \\
\hline
VDM $\rightarrow$ CBIS & 44.8 & {\bf 18.7} & 19.8  \\
CBIS $\rightarrow$ VDM  & 44.8 & 40.4 & {\bf 34.9}  \\
\hline
\end{tabular}
\label{tab:FIDscores}
\end{table}

\begin{table}[ht!]
\caption{Mean (Std Dev) KID scores summary}
\centering
\begin{tabular}{|c|c|c|c|}
\hline
Task & Baseline & cycleGAN & SynDiff
\\
\hline
CMMD $\rightarrow$ CBIS & 0.0220 (0.0047) & 0.0219 (0.0055) & 0.0044 (0.0027)  \\
CBIS $\rightarrow$ CMMD  & 0.0220 (0.0047) & 0.0099 (0.0034) & 0.0091 (0.0027) \\
\hline
CMMD $\rightarrow$ VDM & 0.0173 (0.0040) & 0.0079 (0.0040) & 0.0110 (0.0043) \\
VDM $\rightarrow$ CMMD  & 0.0173 (0.0040) & 0.0091 (0.0036) & 0.0074 (0.0031)  \\
\hline
VDM $\rightarrow$ CBIS & 0.0300 (0.0063) & 0.0060 (0.0036) & 0.0080 (0.0035)  \\
CBIS $\rightarrow$ VDM  & 0.0300 (0.0063) & 0.0164 (0.0059) & 0.0157 (0.0057)  \\
\hline
\end{tabular}
\label{tab:KIDscores}
\end{table}

In Table \ref{tab:FIDscores}, all the tasks for CycleGAN and SynDiff have a FID score lower than the baseline, suggesting that the adapted source images are in greater alignment with the target domain. For the KID (reported in Table \ref{tab:KIDscores}), we pick the number of subsets and subset size to be 50 and 100 respectively. This emulates the resampling of images, i.e., having a high chance that the same sample will be present in multiple bins featured in the original implementation \cite{binkowski2018demystifying}. We did not find any significant changes in the mean KID scores when using higher values of the subset size and numbers of subsets (not reported here). However, we did find some reduction in the standard deviations with larger bin size values. The need to pick an optimal subset size is an unappealing aspect of using the KID. We observe some differences between the mean KID and FID scores. For the VDM $\rightarrow$ CMMD task, the mean KID score for CycleGAN is higher when compared to SynDiff, but Table \ref{tab:FIDscores} shows that the FID score for CycleGAN is lower. It is also interesting to note that the baseline for CMMD $\leftrightarrow$ VDM is the lowest in Table \ref{tab:KIDscores} when compared with the other tasks, but in Table \ref{tab:FIDscores} the baseline is higher than CMMD $\leftrightarrow$ CBIS. Even though these differences could be used to draw opposing conclusions about model performance, the large standard deviation in proportion to the mean KID scores make comparisons challenging. 

\subsection{Numerical precision and the FID}
\label{sec:numerical_prec}
Another factor to consider when using a given metric is whether the numerical precision of the data or data type is appropriate for its use. For example, the PyTorch lightning documentation for FID \footnote{https://lightning.ai/docs/torchmetrics/stable/image/frechet\_inception\_distance.html} states that the metric is known to be unstable in its calculations and that float64 precision is recommended for best results. 
However, we found no differences in the scores computed with float64 precision (Table \ref{tab:FIDscores}) compared to using float32 (not shown here). Additionally, we found that computing the results with float64 precision with our computational resources required approximately twice the time to generate a score compared to using float32.  This suggests that this particular recommendation may not be significant for mammography-based style transfer tasks when using thousands of 256$\times$256 pixel images.

\section{Conclusions}
We have discussed the key challenges with evaluating the performance of unpaired image-to-image translation models, describing the advantages and disadvantages of several popular metrics. We suggest that a more comprehensive assessment may be acquired by using several metrics, each of which consider unique aspects of image quality. We have also emphasised the need to consider how artefacts or other undesirable model behaviours as well as the post-processing steps used to correct them may affect the use/interpretation of common performance evaluation metrics. Ultimately, this work provides a useful reference for others interested in using common metrics to evaluate the performance of generative models, and a starting point for more in-depth analysis.

\addtolength{\textheight}{-12cm}   




\section*{ACKNOWLEDGMENT}
The project 22HLT05 MAIBAI has received funding from the European Partnership on Metrology, co-financed from the European Union’s Horizon Europe Research and Innovation Programme and by the Participating States. Funding for the UK partners was provided by Innovate UK under the Horizon Europe Guarantee Extension.


\bibliographystyle{unsrt}
\bibliography{references}

\end{document}